\documentclass[floatfix,aps,prd,showpacs,amsmath,nofootinbib,preprintnumbers,twocolumn]{revtex4}
\usepackage[latin9]{inputenc}
\usepackage{graphicx}
\usepackage{subfigure}
\usepackage{colordvi}

\def\be{\begin{equation}}
\def\ee{\end{equation}}
\def\ba{\begin{eqnarray}}
\def\ea{\end{eqnarray}}

\begin{document}
\title{Is a Dissipative Regime During Inflation in Agreement with 
Observations?}
\author {Alessandro Cerioni}\email{cerioni@iasfbo.inaf.it}
\affiliation{Dipartimento di Fisica, Universit\`a degli Studi di Bologna and
  INFN via Irnerio, 46-40126, Bologna, Italy}
\affiliation{INAF/IASF-BO,
Istituto di Astrofisica Spaziale e Fisica
Cosmica di Bologna \\
via Gobetti 101, I-40129 Bologna - Italy}
\author{Fabio Finelli}\email{finelli@iasfbo.inaf.it}
\affiliation{INAF/IASF-BO,
Istituto di Astrofisica Spaziale e Fisica
Cosmica di Bologna \\
via Gobetti 101, I-40129 Bologna - Italy}
\affiliation{INAF/OAB, Osservatorio Astronomico di Bologna,
via Ranzani 1, I-40127 Bologna -
Italy}
\affiliation{INFN, Sezione di Bologna,
Via Irnerio 46, I-40126 Bologna, Italy}
\author{Alessandro Gruppuso}\email{gruppuso@iasfbo.inaf.it}
\affiliation{INAF/IASF-BO,
Istituto di Astrofisica Spaziale e Fisica
Cosmica di Bologna \\
via Gobetti 101, I-40129 Bologna - Italy}
\affiliation{INFN, Sezione di Bologna,
Via Irnerio 46, I-40126 Bologna, Italy}
\date{\today}

\begin{abstract}
We study the spectral index of curvature perturbations for  
inflationary models where the driving scalar field is coupled to a 
relativistic fluid through a friction term $\Gamma$. 
We find that only a very small friction term - $\Gamma \ll H$, with H being the 
Hubble parameter during inflation - is allowed 
by observations, otherwise curvature fluctuations are generated 
with a spectral index $n_s$ unacceptably red. 
These results are generic with respect to the inflationary potential 
and known dependence of the friction term on the scalar field and the 
energy density of the relativistic fluid. 
We compare our findings with previous investigations.
\end{abstract}

\maketitle


{\em Introduction.} The inflationary paradigm provides an explanation for the
large-scale curvature fluctuations seen in the pattern of anisotropies of
the cosmic microwave background and of the large scale structure
\cite{books}. Single scalar field inflationary models are successfull 
in matching observations 
if the effective mass of the inflaton - the second derivative of 
its potential $V(\phi)$ - is small compared to the energy scale at which 
inflation occurs \cite{komatsu}. 
Primordial curvature perturbations are then 
generated with a spectrum close to scale invariance by the geometric 
amplification of zero point fluctuations.
When several scalar fields conspire in driving inflation, beyond 
having isocurvature perturbation 
which are constrained by data if survived until 
decoupling between matter and radiation \cite{dunkley}, their mutual interaction 
can drastically change the slope of the primordial spectrum of curvature 
perturbations during the exit from the Hubble radius \cite{KP}.  

{\em Warm Inflation} 
Warm inflation \cite{WI} (see \cite{berera_review} for a review) 
has been proposed as an alternative scenario 
where quantum fluctuations are not amplified just by geometry, 
but also by a production of radiation resulting from a decay of the 
inflaton occuring during the accelerated stage. 
In warm inflation the inflaton - taken in the literature and here as 
a standard scalar field - 
decays into radiation through a friction term in a phenomenological way, 
as in the old theory of reheating \cite{ASTW,DL,AFW}.
The value of the friction term $\Gamma$ 
was originally proposed to be much larger than the Hubble 
rate, but it may be also smaller \cite{BGB}. 
Isocurvature perturbations are generic in this scenario and 
have been studied \cite{TB,LF,DFG},
paying also attention to the post-inflationary evolution \cite{DFG}.
Both for large and small friction terms 
the spectrum of curvature perturbations is nearly scale invariant 
\cite{WI,HMB,BGB,Hall-Peiris}. Concerning the aspect of inflaton interactions, 
the case of warm inflation is therefore at odd with the 
argument above involving scalar fields 
in which the interactions of the inflaton have an important 
impact on the spectral index of curvature perturbations: although 
the key point of the warm inflationary scenario is the interaction of the 
inflaton with a relativistic fluid, no matter of how large this 
interaction is, the resulting spectrum of curvature perturbations is 
claimed to be nearly 
scale invariant in literature, and the strength of the friction term enter just in the small 
deviation from scale invariance also proportional 
to slow-roll parameters \cite{Hall-Peiris}. In this paper we address this 
issue by the use of the standard theory of cosmological perturbations 
considering two different inflationary potential, i.e. a large field model:
\be
V_L (\phi) = \frac{m^2}{2}\phi^2 \,,
\ee
and as small field model, the SUSY breaking potential \cite{berera_PLB}:
\be\label{eq:Hall-Peiris-potential}
V_S (\phi)=\frac{1}{2}\mu^2 \left[ {\phi^2 \log \frac{\phi^2}{\phi_0^2}
+ \phi_0^2 -\phi^2} \right] \,.
\ee
For this last potential, we restrict ourselves to initial conditions for 
the inflaton close to the origin.

{\em Background Eqs}
The simple equations of motion 
for the scalar field and the fluid energy-density $\rho_F$ 
we consider are the following:
\be
\ddot{\phi} = - 3 H \dot\phi - \Gamma \dot\phi - \frac{d V (\phi)}{d \phi}
\label{motophi}
\ee
\be
\dot\rho_F = - 3 H (1 + \omega_F)\rho_F + \Gamma \dot\phi^2\, ,
\label{motoF}
\ee
where the dot denotes the derivative with respect to the cosmic time, 
$\omega_F = p_F/\rho_F$ is the fluid state parameter and the Friedmann 
equation is:
\be
H^2 = \frac{1}{3 M_{\rm pl}^2} \left[ \frac{{\dot \phi}^2}{2} + 
V(\phi) + \rho_F \right] \,, 
\label{hubble}
\ee
with $M_{\rm pl}=1/\sqrt{8\pi G}$ as the reduced Planck mass.
The coupling in Eqs. (3-4) admits a fully covariant 
description \cite{LF,expl}. 
The main idea of warm 
inflation is that dissipation of the inflaton in the perfect fluid 
elonges the slow-roll stage, as can be seen by:
\be
(3 H + \Gamma ) \dot \phi \simeq - \frac{d V (\phi)}{d \phi}
\ee
and gently 
modifies it in a radiation dominated era through a continuous release of 
entropy. The fluid does not redshift as $1/a^{3 (1+w_F)}$ during inflation, 
but is also
almost constant in time ($3 H \rho_F (1+w_F) \simeq \Gamma {\dot \phi}^2$).
By taking as an example a massive inflaton and a decay rate constant in time, 
the elongation of the duration of inflation is 
encoded in the number of e-folds formula: 
\be
N = \frac{\phi_\mathrm{i}^2}{4M_\mathrm{Pl}^2}
\Bigg({1-\frac{\phi_\mathrm{f}^2}{\phi_\mathrm{i}^2}}\Bigg)
+\frac{\Gamma\,\phi_\mathrm{i}}{\sqrt{6}\,M_\mathrm{Pl} \,m}
\Bigg(1-\frac{\phi_\mathrm{f}}{\phi_\mathrm{i}}\Bigg) \,.
\ee
For future convenience we introduce the following adimensional parameter:
\be
\gamma \equiv \frac{\Gamma (\phi, \rho_F) }{3H}\mbox{,}
\ee
and the slow-roll parameters:
\be
\epsilon \equiv -\frac{\dot H}{H^2}\mbox{,}\quad 
\eta_{\phi \phi} \equiv \frac{V_{\phi\phi}}{3H^2}\mbox{,}
\ee
where $V_{\phi\phi}=\frac{d^2V(\phi)}{d\phi^2}.$
We consider the most general case of $\Gamma=\Gamma(\phi,\rho_F)$ 
\cite{HMB}:
\be
\Gamma = \Gamma_0 \left({\frac{\phi}{\phi_0}}\right)^b 
\left({\frac{T}{M}}\right)^c\mbox{,}
\label{gamma_dependence}
\ee
where $M$ is a mass scale, 
$T= (\rho_F/\xi)^{1/4}$ with $\xi \equiv \frac{g_*}{30}\pi^2$, 
$g_*$ being the number of effective relativistic degrees of freedom. 
%

{\em Cosmological Perturbations in Warm Inflation}
For the analysis of scalar cosmological perturbations we follow 
Ref. \cite{DFG} (the set of equations agrees with \cite{LF,HMB}).
In this work the full set of first-order 
equations were given in the uniform curvature gauge (UCG) for scalar 
metric perturbations:
\be
ds^2=-(1+2 \alpha)dt^2 - a \beta_{,i} dt dx^i+a^2 \delta_{ij} dx^i dx^j \,.
\label{PER_UCG2}
\ee
For the fluid the evolution of a redefinition of the 
3-momentum scalar potential $\psi_F$ 
($\delta T^0_{(F)i}\equiv \partial _i \psi_F$) 
\cite{DFG}:
\be
Q_F \equiv \frac{\psi_F}{\sqrt{\rho_F (1 + w_F)}} \,\mbox{,}
\ee
was considered in addition to the scalar field fluctuation $Q_\phi$. 
In the UCG the gauge-invariant
scalar field and fluid fluctuations used for canonical quantization
\cite{mukhanov,lukash} coincide with the field fluctuation
$Q_\phi$ and $Q_F$.
In the UCG the gauge-invariant 
comoving curvature perturbation is then simply written as the weighted sum 
of the field fluctuation $Q_\phi$ and the 3-momentum scalar potential of 
the fluid:
\be
\mathcal{R} = H
\frac{\dot \phi \, Q_\phi - \sqrt{\rho_F (1 + w_F)} \, Q_F}{{\dot \phi}^2
+ \rho_F ( 1 + w_F) } \mbox{.}
\ee
in full analogy with the more studied case of the two scalar field 
inflationary model system \cite{FB}:
\be
\mathcal{R} = H
\frac{\dot \phi \, Q_\phi + \dot \chi \, Q_\chi}{{\dot \phi}^2
+ {\dot \chi}^2} \, \mbox{.}
\ee
Field and fluid fluctuations evolve in time according to \cite{DFG}:
\be
{\ddot Q_i} + \left( 3 H \, \delta_{i j} + G_{i j} \right) \, {\dot Q_j} +
{\Omega_{i j}} \, Q_j =0
\label{system}
\ee
where $(i \,, j) = (\phi \,, F)$ and $G_{i j} \,, \Omega_{i j}$ are given 
in Ref. \cite{DFG} for $\Gamma={\rm const.}$ and generalized to 
$\Gamma = \Gamma (\phi, \rho_F)$ in \cite{tesi,talk}.
We have run numerical simulations following 
the system in Eq. (\ref{system}) with 
initial conditions:
\be\label{eq:initial-conditions}
Q_{\phi|k} \simeq\frac{e^{-ik(\tau-\tau_i)}}
{a^{1+\frac{3}{2} \gamma}\sqrt{2k}}\mbox{,}\qquad Q_{F|k}
\simeq\frac{e^{-i\omega_F^{1/2}k(\tau-\tau_i)}}{a^{1+\frac{3}{2}(1+\omega_F)}
\sqrt[4]{4k^2\omega_F}}\mbox{,}
\ee
when $\tau$ stands for the conformal time coordinate 
(and $\tau_i$ is some initial time). 
These have been obtained analytically solving the evolution equations 
for $Q_{\phi|k}$ 
and $Q_{F|k}$ (shown in Ref. \cite{DFG}) with $k\gg aH$ and 
neglecting self-consistently cross terms in the time derivatives; 
the normalization factor has been evaluated in order to match the 
plane wave solution at early times.
The power spectrum for the gauge-invariant comoving curvature perturbation 
\be
\mathcal{P}_\mathcal{R}=\frac{k^3 |\mathcal{R}_k|^2}{2\pi^2}=
A_s\left({\frac{k}{k_*}}\right)^{n_S-1}\mbox{,}
\ee
is evaluated numerically at the end of inflation. $k_*$ is a pivot scale, 
which is chosen to exit from the Hubble radius $51$ e-folds from the end 
of inflation; we consider a range of modes which spans three orders of 
magnitude centered around the pivot scale $k_*$.

{\em Numerical Results}
The results for 
$n_S -1 $ are presented in 
Table I and II for $w_F=1/3$, for the large and small field model, respectively. 
For the large field model only the case 
$\Gamma = {\rm const.}$ ($b=c=0$ in Eq. (\ref{gamma_dependence})) 
is considered and one can see that already $\Gamma \sim m$ is inconsistent 
with observations, given the last constraints on the spectral index 
\cite{komatsu,dunkley}.
For the small field model several possibilities allowed 
by Eq. (\ref{gamma_dependence}) are considered with 
$M^2=\mu \phi_0/\sqrt{2} \,, g_*=100$: 
again $\gamma_* \gtrsim 0.01$ is inconsistent with observations. 

{\em Analytic argument for red spectra}
The numerical results can be easily understood analitically by 
considering the equation for $Q_\phi$ 
without potential and without coupling to $Q_F \,, \dot Q_F$ 
in a de Sitter space-time with $\Gamma$ constant in time. 
The equation for this massless fluctuation admits 
exact solutions in terms of the Hankel functions \cite{AS}:
\be
Q_{\phi|k} = \sqrt{\frac{\pi}{4 H}} 
\frac{H_\nu^{(1)}(-k\eta)}{a^{3(1+\gamma)/2}}
\ee
with $\nu = \frac{3}{2} + \frac{\gamma}{2}$. This solution concides for 
$\gamma=0$ with the massless solution in de Sitter and 
has a solution constant in time for $-k \eta <<1$ 
with a spectral index $n_S-1 = -3 \gamma$.
Following \cite{Stewart:1993bc}, we provide also an expression 
for a quasi-De Sitter background in the simplest case of 
$\Gamma =\mathrm{const.}$: 
\be\label{analytic-spectral-index}
n_S-1\simeq-3\,\gamma_*+\frac{2}{1+\gamma_*}\,\eta_*-6\frac{1+\gamma_*
+\frac{1}{2}\gamma_*^2}{1+\gamma_*}\,\epsilon_*\mbox{,}
\ee
where $*$ stands for the time at
which the pivot scale crosses the Hubble radius ($k_*=a_*\,H_*$).
Eq. (\ref{analytic-spectral-index}) agrees with the previous de Sitter 
expression for $\epsilon=\eta_{\phi \phi}=0$ and with the standard results of cold inflation \cite{Stewart:1993bc} 
for $\gamma=0$. Our expression does not agree with the result quoted in literature for warm inflation in the 
quantum regime \cite{BGB,Hall-Peiris}: 
\be
n_s - 1 |_{\rm literature} = 6\epsilon_*+2\eta_*+
\gamma_*\left(8\epsilon_*-2\eta_*-2\beta_*\right)
\label{literature}
\ee
where $\beta = M_{\rm pl}^2 \Gamma_\phi V_\phi/(\Gamma V)$, 
but agrees very well with our numerical results \cite{diffs}, as can be 
seen from Table I and II. Note how our result in 
Eq. (\ref{analytic-spectral-index}) contains the 
dependence on $\Gamma$ linearly and not multiplied by the slow-roll 
parameters as in Eq. (\ref{literature}): 
the simple intuitive and correct argument to explain  
our result is the occurrence of $\Gamma$ in the equation of motion for the 
scalar field and its fluctuations not weighted by any slow-roll parameter.


\begin{table}
\begin{centering}
\begin{footnotesize}
\begin{tabular}{|c|c|c|c|c|c|c|}
\hline 
{\small $\frac{\Gamma_0}{m}$} & {\small 
$\frac{\phi_{\mathrm{i}}}{m_{\mathrm{Pl}}}$ } & {\small 
$\frac{H_{\mathrm{i}}}{m}$ } & $\gamma_*$ 
& {\small $n_S-1$ } & {\small $n_S-1$ } & 
{\small $n_S-1$ }  
\tabularnewline
{} & {} & {} & {} & { numerical } & { analytical } & {literature} \tabularnewline
\hline
\hline 
0.01 & 3.34 & 6.83 & 0.00059 &  -0.0433 &  -0.0433  & -0.0415 \tabularnewline
1    & 3.18 & 6.51 & 0.061 &  -0.219    & -0.223  &  -0.364  \tabularnewline
8    & 2.28 & 4.67 & 0.71 & -2.17 & -2.21 & 0.00965 \tabularnewline
\hline
\end{tabular}
\end{footnotesize}
\end{centering}
\caption{Numerical and analytical spectral indexes in case of quadratic 
potential and $\Gamma=\mathrm{const.}$; the 
i subscript stands for 
initial and $*$ stands for the time at 
which the pivot scale crosses the Hubble radius ($k_*=a_*\,H_*$).}
\label{tab:spectral-indexes-large-field}
\end{table}

\begin{table}
\begin{centering}
\begin{footnotesize}
\begin{tabular}{|c|c|c|c|c|c|c|c|c|c|}
\hline 
{\small $\frac{\Gamma_0}{\mu}$} & {\small $b$} & {\small $c$} & 
{\small $\frac{\phi_i}{\phi_0}$ } & {\small $\frac{H_{\mathrm{i}}}{\mu}$ } 
& $\gamma_*$ & 
{\small $n_S-1$ } 
& {\small $n_S-1$ }& {\small $n_S-1$ }
\tabularnewline
{} & {} & {} & &  & {} & { numerical } & { analytical }   & {literature} \tabularnewline
\hline
\hline 
0.01 & 1  &  0 & 0.37 & 7.87 & 0.00022 &  -0.0398 & -0.403  & -0.0395 \tabularnewline
1    & 1  &  0 & 0.39 & 7.67 & 0.023 &   -0.109   &  -0.111 & -0.0268  \tabularnewline
8    & 1  &  0 & 0.49 & 6.66 & 0.28 &  -0.854 & -0.880 &  0.133  \tabularnewline
\hline
0.01 & 2  &  0 & 0.37 &   7.87 & 0.00010 & - 0.0388 &  -0.0399 & -0.0395  \tabularnewline
1    & 2  &  0 & 0.38 &  7.77 & 0.011 &  -0.0713 &  -0.0734 & -0.0284  \tabularnewline
8    & 2  &  0 & 0.46 &  6.99 & 0.14  &  -0.461  &  -0.471   & 0.120     \tabularnewline
\hline
0.01 & 1  & -1 & 0.38 & 7.77 & 0.0086 & -0.0603 & -0.0658  & -0.347 \tabularnewline
0.1   & 1  & -1 & 0.40 & 7.57 & 0.059    &  -0.201  & -0.219  & -0.00544 \tabularnewline
1      & 1  & -1 & 0.59 & 5.46 & 0.83    & -2.48  & -2.56  &    0.591 \tabularnewline
\hline
0.01 & 2  & -1 & 0.37 & 7.87 & 0.0047 &  -0.0474 & -0.0539 &  -0.0346  \tabularnewline
0.1   & 2  & -1 & 0.38 & 7.67 & 0.032 &  -0.135 & -0.138 &  -0.00457   \tabularnewline
1      & 2  & -1 & 0.56 & 5.81 & 0.50 & -1.54  &  -1.57 & 0.644 \tabularnewline

\hline
\end{tabular}
\end{footnotesize}
\par\end{centering}

\caption{Numerical and analytical spectral indexes in case of 
small field model; the
i subscript stands for initial and $*$ stands for the time at 
which the pivot scale crosses the Hubble radius ($k_*=a_*\,H_*$). 
}
\label{tab:spectral-indexes-with-field-dependence}

\end{table}

\begin{figure*}
\begin{tabular}{cc}
\subfigure[\ Large field model, with $\Gamma=8\,m$.]{
\includegraphics[width=0.23\textwidth]{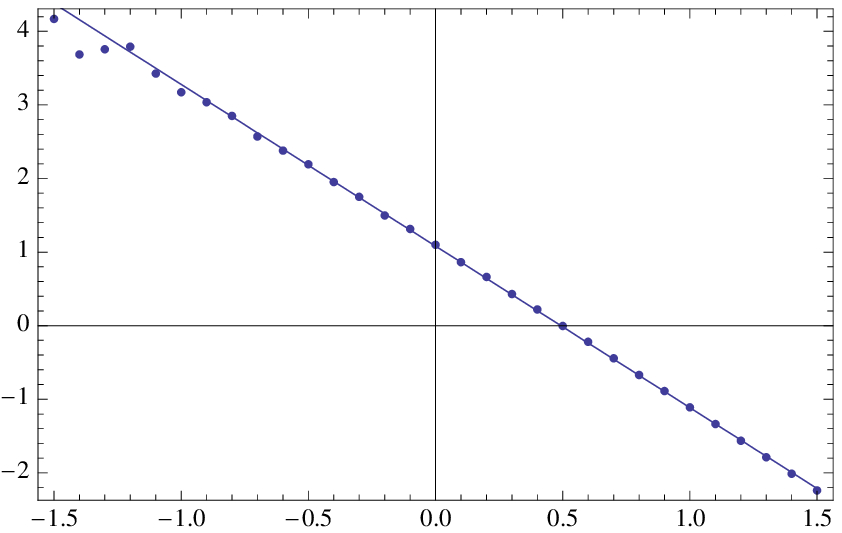}}
\subfigure[\ Small field model, with $\Gamma_0 = 1\,\mu$, $b=1$, $c=0$.]{
\includegraphics[width=0.23\textwidth]{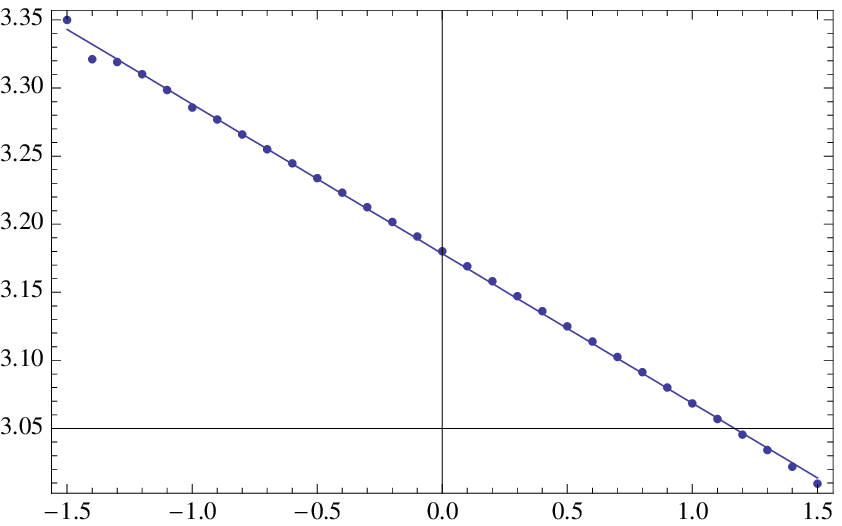}}
\subfigure[\ Small field model, with $\Gamma_0 = 1\,\mu$, $b=2$, $c=0$.]{
\includegraphics[width=0.23\textwidth]{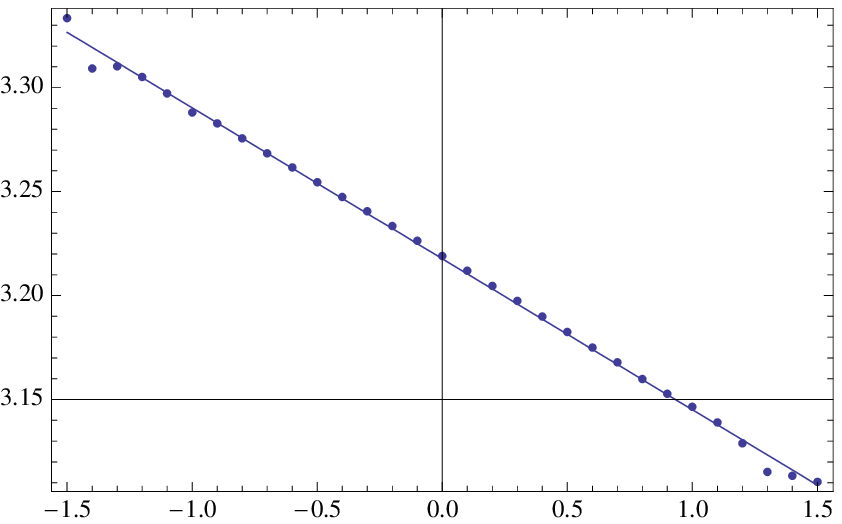}}
\subfigure[\ Small field model, with $\Gamma_0 = 1\,\mu$, $b=2$, $c=-1$.]{
\includegraphics[width=0.23\textwidth]{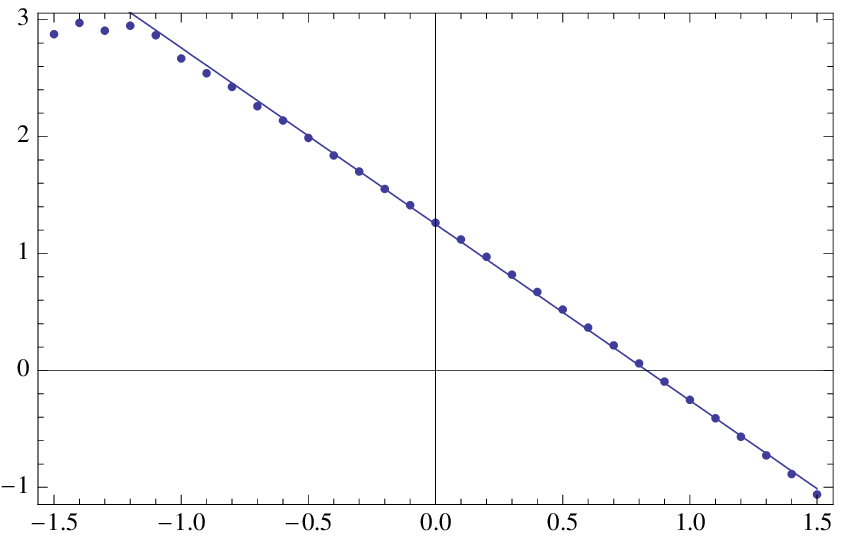}}
\end{tabular}
\caption{Logarithm of the curvature perturbations power spectra 
vs $\log_{10} (k/k_*)$. 
The dots are the spectra of curvature perturbations 
numerically evaluated at the freeze-out 
scale $k_* = a_* \sqrt{\Gamma_* H_*}$ in four case, one for the large 
field case and three for the small 
field case. The solid lines are the predictions for curvature perturbations 
based on Eq. (\ref{analytic-spectral-index}).
Note how the spectra are far from being scale invariant, which is 
the key prediction of Ref. \cite{HMB}, and almost identical to their 
long-wavelength spectra.
}
\label{fig1}
\end{figure*}

{\em Explanation of the Discrepance}
A difference between our study and Ref. (\cite{HMB}) 
in the treatment of perturbations when they are in the sub-Hubble regime. 
Our treatment is 
the standard one which takes into account metric fluctuations 
self-consistently and which 
is used for any computation in a universe dominated by two scalar fields or 
fluids. Ref. \cite{HMB} uses in this regime a Langevin equation for the 
field fluctuations in rigid space-time driven by a thermal bath claiming that 
fluctuations are scale invariant at freeze-out, i.e. at 
$k=a\,\sqrt{\Gamma H}$.
Fig. (\ref{fig1}) shows how the spectrum of curvature perturbations 
when evolved according to the
system in Eq. (\ref{system}) with initial conditions 
(\ref{eq:initial-conditions}) is already red tilted 
at the freeze-out scale and far from scale invariance. 
We therefore conclude that the Langevin approximation used 
for sub-Hubble scales in \cite{HMB} does not agree with the standard 
theory of cosmological perturbations.

{\em Conclusions}
Warm inflation is an alternative to the pure 
geometric amplification in producing a nearly scale-invariant spectrum of 
curvature perturbations. It was argued \cite{YL} 
that it is difficult to realize a strong dissipative regime described 
by Eqs. (3-5) from quantum field theory at finite
temperature. In this paper we have shown a standard treatment of cosmological perturbations for 
the background system in Eqs. (3-5) leads to a
spectrum of curvature perturbations which agrees with observations only when the dissipation 
is weak, starting from quantum initial fluctuations. It is not clear if the 
linear dependence of the spectral index of curvature 
perturbations on $\gamma$ found here may change by starting from thermal 
initial conditions, rather than from quantum ones, since quantum 
fluctuations are present anyway.
It is however interesting that an introduction of a coupling along the 
lines of Eqs. (3-5) may bring cold inflationary models  
predicting a blue spectrum back in agreement with current observations.

\acknowledgments

We wish to thank A. Berera, R. Brandenberger, L. Hall, A. Linde and 
H. Peiris for comments and discussions on this project.
We thank INFN IS BO11 for partial support.
F. F. is partially supported by INFN IS PD 51.



\begin{thebibliography}{99}
\bibitem{expl}
In covariant form the conservation equation for the fluid are
written as $\nabla_\mu T^\mu_{F \,, \nu} = -
\Gamma (\partial_\beta \phi) u^\beta \partial_\nu \phi$, with
$u_\beta$ being a normalized time-like vector.
\bibitem{diffs}
The percent difference between the numerical result and 
our analytic expression in Eq. (\ref{analytic-spectral-index}) can be due 
to a) the way we approximate in 
closing the dynamics of $Q_\phi$ in one single equation during inflation 
and/or  
b) the presence of isocurvature modes feeding the curvature perturbation and 
modifying slightly its spectral index during and after inflation and/or 
c) the deviations from $\Gamma = {\rm const.}$ when relevant.
\end{thebibliography}
\end{document}